\documentclass[usenatbib]{mn2e}
\usepackage{amssymb}

\usepackage{graphicx}
 
\title[Binary nature of AGL~J2241+4454 and HESS~J0632+057]
{On the binary nature of the $\gamma$-ray sources 
AGL~J2241+4454 (=MWC~656) and HESS~J0632+057 (=MWC~148)}

\author[J. Casares et al.]
{J. Casares$^{1,2}$\thanks{E-mail: jorge.casares@iac.es (JCV); mribo@am.ub.es (MR);  
iribas@ieec.uab.es (IR); jmparedes@ub.edu (JMP); vilardell@ice.cat (FV); 
ignacio@dfists.ua.es (IN)}  
 M. Rib\'o$^3$, I. Ribas$^{4}$, J.M. Paredes$^3$, F. Vilardell$^{5}$, I. Negueruela$^5$\\
$^1$ Instituto de Astrof\'{\i}sica de Canarias, E-38200 La Laguna, Tenerife, Spain\\
$^2$ Departamento de Astrof\'{\i}sica, Universidad de La Laguna, Avda.
Astrof\'{\i}sico Francisco S\'anchez s/n, E-38271 La Laguna, Tenerife,\\
~~~Spain\\
$^3$ Departament d'Astronomia i Meteorologia, Institut de 
Ci\`encies del Cosmos (ICC), Universitat de Barcelona (IEEC-UB), \\
~~~Mart\'\i{} i Franqu\`es 1, E-08028 Barcelona, Spain\\
$^4$ Institut de Ci\`encies de l'Espai -- (IEEC-CSIC), Campus UAB, Facultat de 
Ci\`encies, Torre C5 - parell - 2a planta, E-08193 \\
~~~Bellaterra, Spain\\
$^5$ Departamento de F\'{\i}sica, Ingenier\'{\i}a de Sistemas y Teor\'{\i}a de la Se\~nal, 
Universidad de Alicante, Apdo. 99, E-03080 Alicante, Spain
}

\begin{document}

\maketitle

\begin{abstract} 
\noindent We present optical spectroscopy of MWC~656 and MWC~148, the proposed 
optical counterparts of the $\gamma$-ray sources AGL J2241+4454 and HESS J0632+057,
respectively.  
The main parameters of the H$\alpha$ emission line ($EW$, $FWHM$ and centroid 
velocity) in these stars are modulated on the proposed orbital periods of 60.37 
and 321 days, respectively. These modulations are likely produced by the resonant 
interaction of the Be discs with compact stars in eccentric orbits. We also present 
radial velocity curves of the optical stars folded on the above periods and obtain 
the first orbital elements of the two $\gamma$-ray sources thus confirming their 
binary nature. Our orbital solution support eccentricities $e\sim0.4$ and 0.83$\pm$0.08 
for MWC~656 and MWC~148, respectively. Further, our orbital elements imply that the 
X-ray outbursts in HESS~J0632+057/MWC~148 are delayed $\sim$0.3 orbital phases after 
periastron passage, similarly to the case of LS~I~+61~303. In addition, the optical 
photometric light curve maxima in AGL~J2241+4454/MWC~656 occur $\sim$0.25 phases 
passed periastron, similar to what is seen in LS~I~+61~303. We also find that the 
orbital eccentricity is correlated with orbital period for the known $\gamma$-ray binaries. 
This is explained by the fact that small stellar separations are required for the 
efficient triggering of VHE radiation. Another correlation between the $EW$ of 
H$\alpha$ and orbital period is also observed, similarly to the case of Be/X-ray 
binaries. These correlations are useful to provide estimates of the key orbital 
parameters $P_{\rm orb}$ and $e$ from the H$\alpha$ line in future Be $\gamma$-ray 
binary candidates.    

\end{abstract}


\section{Introduction}

Cherenkov telescopes are opening a new era of discoveries with the detection of 
large populations of $\gamma$-ray sources ranging from galactic objects
(supernovae remnants, pulsar-wind nebulae, compact binaries) to distant 
blazars and starburst galaxies \citep{hinton09a}. In particular, the class of 
$\gamma$-ray binaries has attracted great attention in the last few years. 
They are characterised by the presence of a compact star 
orbiting a late-O/early B companion. 
Either if particles are accelerated in a microquasar jet or in the collision of
two winds, the high density of UV photons from the 
massive star provides the necessary environment for the production of GeV-TeV 
emission through inverse Compton scattering. 
There are currently three confirmed $\gamma$-ray binaries with High Energy 
(HE, E$>$100 MeV) and/or Very High Energy (VHE, E$>$100 GeV) emission modulated 
on the orbital period: PSR B1259$-$63, LS 5039 and LS~I~+61~303 \citep{paredes11}. 
Three more candidates have been recently proposed but their binary nature 
awaits confirmation. These are 1FGL J1018.6$-$5856 \citep{corbet11}, AGL 
J2241+4454 and HESS J0632+057. The latter two sources are the subject of this 
paper. 

AGL J2241+4454 is a point-like $\gamma$-ray source detected by the AGILE 
satellite above 100 MeV \citep{lucarelli10}. The error circle of the satellite
contains the Be star MWC~656 (= HD~215227 ) which was proposed by 
\cite{williams10} as the likely optical counterpart. The same authors suggest 
a spectral classification B3~IVne+sh and an orbital period of 60.37$\pm0.04$~d
based on Hipparcos light curve data. By fitting the spectral energy distribution 
in the UV and $B-$band a distance of 2.6 $\pm$ 1.0 kpc is derived. This,
combined with its high Galactic latitude $b=-12^{\circ}$, implies a runaway
star, another evidence supporting its binary nature. 

On the other hand, MWC~148 (=HD~259440) is a 
B0~Vpe star which has been proposed as the optical counterpart of the  
TeV source HESS~J0632+057 \citep{aharonian07}. The claim is 
based on the detection of radio \citep{skilton09} and X-ray emission 
associated to MWC~148 \citep{hinton09b,falcone10} with 
similar spectral indexes and variability as observed in 
PSR~B1259$-$63, LS~5039 and LS~I~+61~303. 
Recently, extended and variable radio emission at AU scales, showing similar 
morphology to these three $\gamma$-ray binaries, has been reported 
(Mold\'on et al. 2011a; see also Mold\'on et al. 2011b, Rib\'o et al. 2008 and 
Dhawan et al. 2006).
Lower limits to the orbital period of $>$54 d \citep{falcone10} and $>$200 d 
\citep{casares11} were proposed based on X-ray and optical emission line 
variability. Further, the absence of significant radial velocity shifts in 
the B0 star supports a long period $>$100 d \citep{aragona10}. 
Finally, \cite{bongiorno11} report the presence of strong X-ray flares in 
{\it Swift}/XRT data, modulated with a period of 321 $\pm$ 5 d which strongly 
advocates for its binary nature.     

Alternative scenarios to the binary hypothesis invoke an isolated 
star where high energy emission is produced in a wind driven shock. In 
particular, Bp stars are known to posses strong surface magnetic fields which 
efficiently confine the stellar 
wind and may accelerate particles up to TeV energies \citep{townsend07}. 
The best way to test the binary model and solve the system geometry is through 
a radial velocity study and 
hence we have embarked in an optical spectroscopic campaign of MWC 656 and MWC
148 using several telescopes. 
Our analysis demonstrates that these stars are indeed the optical companions of 
High Mass X-ray/$\gamma$-ray Binaries responsible for the HE/VHE emission of 
AGL J2241+4454 and HESS J0632+057. 
In the following sections we present the observations, our best constraints to
the orbital parameters and discussion of results.

\begin{table*}
\centering
\caption[]{ Log of the  MWC~656 observations.}
\label{tab1}
\scriptsize
\begin{tabular}{lccccc}
\hline
\hline
Date &  Telescope & Spect. Range & Number of & Exp. Time & Dispersion  \\
 & & \AA\ & spectra &(seconds) &  (\AA\ pix$^{-1}$) \\
\hline
23 Apr -- 28 Jul 2011 & LT & 3900--5215 & 32 & 600  & 0.35 \\
23 Apr -- 28 Jul 2011 & LT & 5900--8000 & 64 & 290  & 0.80 \\
\hline
\end{tabular}
\end{table*}
\begin{table*}
\centering
\caption[]{ Log of the MWC~148 observations.}
\label{tab2}
\scriptsize
\begin{tabular}{lccccc}
\hline
\hline
Date &  Telescope & Spect. Range & Number of & Exp. Time & Dispersion  \\
 & & \AA\ & spectra & (seconds) &  (\AA\ pix$^{-1}$) \\
\hline
20 Oct 2008 & INT & 3900--5500 & 10 & 90  & 0.63 \\
21 Oct 2008 & INT & 3900--5500 & 5 & 90  & 0.63 \\
22 Oct 2008 & INT & 3900--5500 & 10,10 & 120,90 & 0.63 \\
07 Dec 2008 & WHT & 3800--5400 & 3 & 20& 0.88 \\
07 Dec 2008 & WHT & 8300--8960 & 3 & 20& 0.99 \\
08 Dec 2008 & WHT & 3800--5400 & 3 & 20& 0.88 \\
08 Dec 2008 & WHT & 5555--7070 & 3 & 20& 0.99 \\
10 Dec 2008 & WHT & 3700--4500 & 3 & 25& 0.45 \\
10 Dec 2008 & WHT & 8335--9075 & 3 & 15& 0.48 \\
11 Dec 2008 & WHT & 3700--4500 & 3 & 25& 0.45 \\
11 Dec 2008 & WHT & 8335--9075 & 3 & 15& 0.48 \\
30 Oct-8 Nov 2009 & MT & 3770--9000 & 27 & 1800& 0.02 \\
20 Jan-8 Apr 2010 & ST & 3870--8800 & 22 & 1800& 0.10 \\
4 Sep 2010-3 May 2011 & LT & 3900--5215 & 40 & 600  & 0.35 \\
4 Sep 2010-20 Oct 2010 & LT & 5900--8000 & 48 & 193  & 0.80 \\
24 Oct 2010-3 May 2011 & LT & 5900--8000 & 47 & 290  & 0.80 \\
\hline
\end{tabular}
\end{table*}

\section{Observations}

\subsection{MWC~656}

We observed MWC~656 using the Fibre-fed RObotic Dual-beam Optical Spectrograph 
(FRODOspec) on the robotic 2.0m Liverpool telescope (LT) at the Observatorio del 
Roque de Los Muchachos between 23 April and 28 July 2011. The spectrograph is 
fed by a fiber bundle array consisting of 12x12 lenslets of 0.82$\arcsec$ each which is 
reformatted as a slit. The spectrograph was operated in high resolution mode, 
providing a dispersion of 0.35 \AA ~pix$^{-1}$ and spectral resolving power 
$R\sim5500$ in the blue arm while 0.80 \AA ~pix$^{-1}$ and $R\sim5300$ in the 
red arm. The spectral coverage was 3900--5215 \AA ~and 5900--8000 \AA 
~respectively. A total of thirty two 600s spectra were obtained with the 
blue arm and sixty four 290s spectra with the red arm. One blue and two 
red spectra were obtained on each observing night. A log of the
observations is presented in Table \ref{tab1}. 
The FRODOspec pipeline produces fully extracted and wavelength calibrated
spectra with rms$\le$0.1 \AA ~above 4400 \AA. The analysis presented in this
paper has been performed with the FRODOspec pipeline products.

\subsection{MWC~148}

MWC 148 was observed with the Intermediate Dispersion Spectrograph (IDS)
attached to the 2.5~m Isaac Newton Telescope (INT) at the Observatorio del
Roque de Los Muchachos on the nights of 20--23 October 2008. A total of 35
spectra were obtained in the blue spectral range (3900--5500 \AA,
unvignetted) using the R900V grating in combination with the 235~mm
camera and a 1.2$\arcsec$ slit to provide a spectral resolution of 80 km 
s$^{-1}$ (FWHM). Nightly averages were produced from the individual 
spectra resulting in a total of 3 INT spectra. 

Twelve additional blue spectra were obtained on the nights of 7--8 and 10--11 
December 2008 with the Intermediate dispersion Spectrograph and Imaging 
System (ISIS) double-arm spectrograph on the William Herschel Telescope (WHT). 
Here we employed the 600B and 1200B gratings on the blue arm resulting 
in wavelength coverages 3800--5400 \AA ~and 3700--4500 \AA ~respectively.  
A 1$\arcsec$ slit was selected yielding spectral resolutions in the range 65--115 km 
s$^{-1}$. Meanwhile, the 600R and 1200R gratings were mounted on the ISIS red 
arm, centered at different wavelengths, to yield simultaneous spectra in the 
H$\alpha$ and CaII NIR triplet. Nightly averages were produced for the 
different spectral configurations.  

The High Efficiency and Resolution Mercator Echelle Spectrograph (HERMES) on 
the 1.2m MERCATOR (MT) telescope was also used to obtain three nightly 1800 
second spectra on the nights of 30--31 October, 1--4 and 6--8 November 2009. 
HERMES is a fiber-fed spectrograph and we employed the high resolution mode 
which yield a resolving power $R=85000$ accross the entire optical range 
between 3770--9000 \AA. The signal-to-noise in the HERMES spectra is 
significantly lower than in the INT and WHT data so we decided to co-add the 
three individual spectra obtained every night resulting in a total of 9 
Mercator spectra. 

Twenty two high-resolution echelle spectra were also obtained in the period 
Jan 20--Apr 8 2010 using the fiber-fed STELLA Echelle Spectrograph (SES) of 
the 1.2m robotic STELLA-I (ST) telescope at the Observatorio del Teide in 
Tenerife. The spectra cover the wavelength range 3870--8800 \AA ~with increasing
inter-order gaps starting at 7200 \AA. The spectrograph provides an effective
resolving power $R=$55000. One spectrum was obtained per night and the exposure 
time was set to 1800 s.  
 
Finally, MWC~148 was also observed with FRODOspec at the LT between 4 Sep 2010
and 3 May 2011 using the same configuration as for MWC~656 (see above). 
One 600s spectrum and two 290 s or three 193 s red spectra were obtained per 
night. A full log of the observations is presented in Table \ref{tab2}.   

The INT, WHT and MT spectra were reduced using standard techniques  
including debiasing and flatfielding. The spectra were subsequently extracted 
using optimal extraction techniques in order to optimize the signal-to-noise 
ratio of the output \citep{horne86}.  
Frequent observations of comparison arc lamp or hollow cathode lamp images
were performed in the course of each run and the pixel-to-wavelength scale
was derived through polynomial fits to a large number of identified
reference lines. The final rms scatter of the fit was always $<$1/30 of
the spectral dispersion. The automatic pipeline products were used for the 
ST and the LT data.

The spectral type standard HR 2479 (B0 III) was also observed with all 
different telescopes and instrument configurations for the purpose of computing the 
rotational broadening of the companion star. It was selected because of its 
low projected rotational velocity of 50 km s$^{-1}$\citep{abt02}. 

\begin{figure}
\centering
\includegraphics[angle=0,width=90mm]{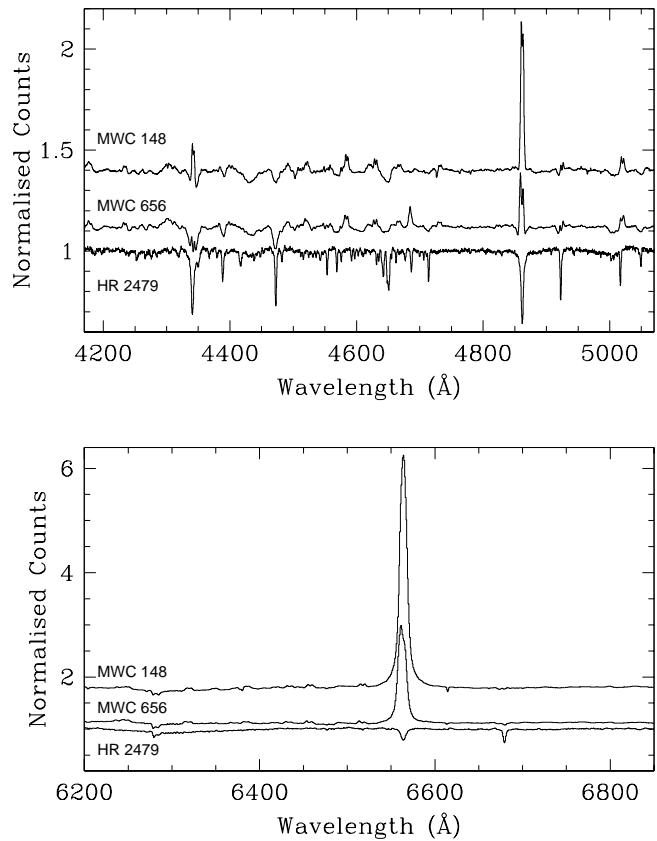}
\caption{Average LT spectra of MWC 148 and MWC 656 in the blue (top panel) and 
red (bottom panel) spectral regions. 
The bottom spectra show the B0 III star HR~2479 for comparison, which has an 
intrinsic broadening of 50 km s$^{-1}$.  
}
\label{fig1}
\end{figure}

\section{Description of the spectra and rotational broadening}

All the spectra were rectified by dividing a low order spline fit to the 
continuum. For comparison, Figure~\ref{fig1} displays the averaged normalised 
LT spectra of MWC~148 and MWC~656. Both targets show similar spectra, with very 
broad and shallow photospheric absorption lines. Strong Balmer emission is seen 
out to H$\beta$ in MWC~656 and to H$\gamma$ in MWC~148. The H${\beta}$ line 
shows a double-peaked profile characteristic of circumstellar discs. Several 
Fe~{\sc ii} disc emission lines are also detected. In summary, both spectra are 
typical of early-type Be stars with well developed discs fed by stable mass 
loss \citep{porriv03}. 
Detailed spectral analysis performed by \cite{williams10} and \cite{aragona10} 
support a spectral classification B3 IVne+sh for MWC~656 and B0 Vpe  for 
MWC~148. In the remaining of this paper we have adopted the spectral 
classification and stellar parameters derived in these works.

Projected rotational velocities of 430 km 
s$^{-1}$ \citep{guti07} and 500 km s$^{-1}$ \citep{aragona10} have been 
reported for MWC~148 whereas 262$\pm$26 km s$^{-1}$ \citep{yudin01} and 300$\pm$50 km 
s$^{-1}$ \citep{williams10} for MWC~656.    
We decided to estimate the rotational broadening $v \sin i$ from our own spectra 
following the technique outlined in \cite{marsh94}, which 
basically subtracts broadened templates from the average spectrum of the target  
and searchs for the lowest residual. Therefore, we first rebinned our LT 
spectra of the two stars and the template HR 2479 into a uniform velocity scale 
of 23 km s$^{-1}$ per pixel. The template was subsequently broadened 
from 100 to 500 km s$^{-1}$ in steps on 5 km s$^{-1}$ using 
a Gray rotational profile \citep{gray92} with a limb darkening coefficient 
$\epsilon=0.33$. 
The broadened versions of the template were multiplied by a scaling
factor $\le1$ to account for dillution due to extra sources of continuum light
(such as the equatorial Be disc) and subtracted from the averaged 
spectra of the two targets. The subtraction was restricted to the blue 
spectral range 4400--5200 \AA, where the photospheric He~{\sc i}, He~{\sc ii} 
and metallic lines are most prominent. The IS, Balmer and 
Fe~{\sc ii} emission lines were all masked in the process. 
A $\chi^2$ test on the residuals of the subtraction yields optimum 
rotational broadening of 370 km s$^{-1}$ for MWC~148 and 342 km s$^{-1}$ for
MWC~656, with a formal uncertainty of $\pm$ 10 km s$^{-1}$. The rotational
velocity of the template star 
needs to be added quadratically
to these, resulting in $v \sin i=373$ km s$^{-1}$ for MWC~148 and 346 km s$^{-1}$ 
for MWC~656. The same analysis was repeated for MWC~148 using the other 
instrument configurations and we always obtain values in the range 355--381 km 
s$^{-1}$. 

\begin{figure*}[ht!]
\centering
\includegraphics[angle=0,width=175mm]{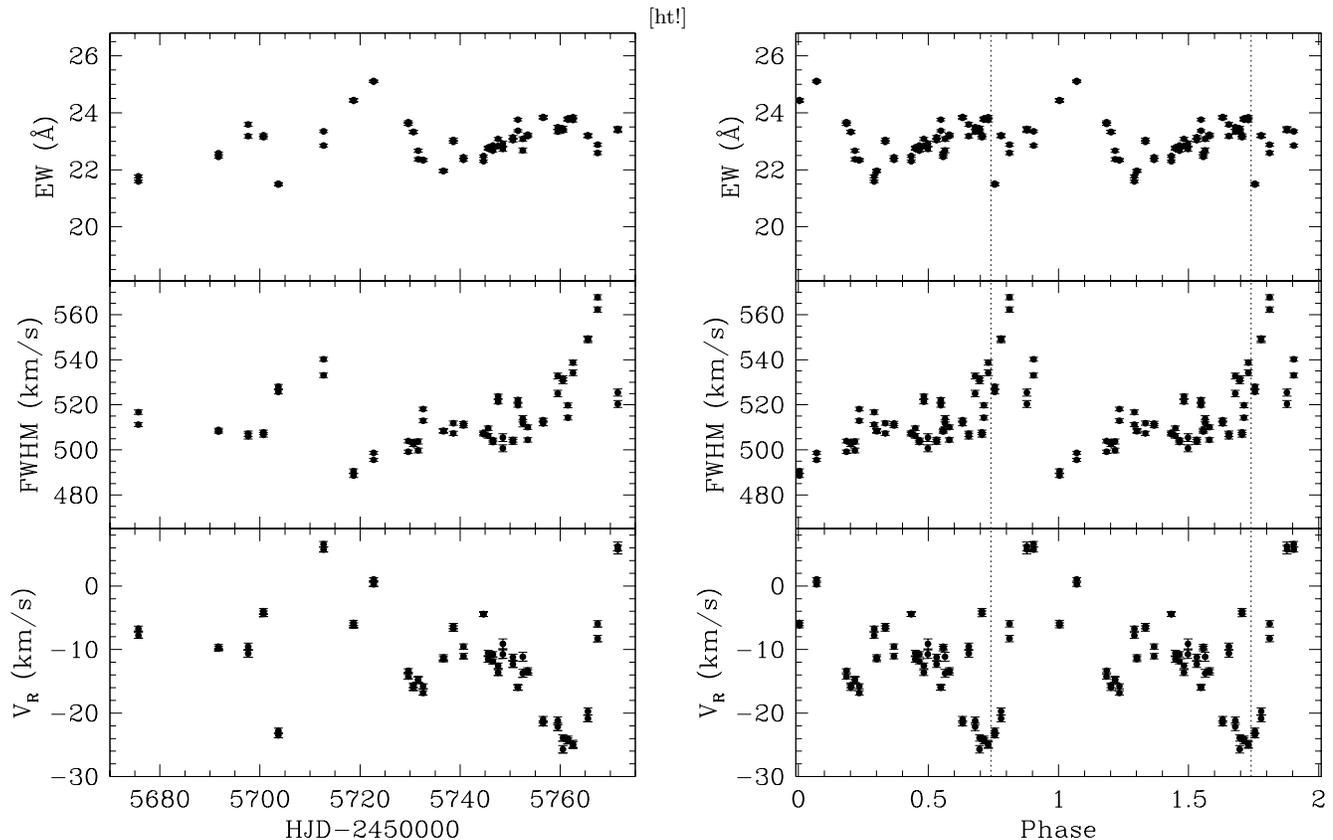}
\caption{Left: time evolution of the H$\alpha$ line parameters in MWC~656. Both the
$FWHM$ and centroid velocities show significant variability with maxima 
roughly separated by $\sim$60 d. Right: Same but folded on the 60.37 d period of 
Williams et al. (2010). Phase 0 has been set to the time of maximum optical brightness  
at HJD 2453243.3. The vertical dotted line denotes the phase of the periastron 
(see text).}
\label{fig2}
\end{figure*}

\section{Analysis of MWC~656}

Be discs in X-ray binaries typically display long-term superorbital variability 
associated to 
activity episodes, but sometimes also variability modulated with the binary 
orbital period \citep{zamanov99}. The strong H$\alpha$ emision is the best 
tracer of disc variabiliy and hence we decided to analyse several line 
parameters in MWC~656 and test whether they are modulated with the claimed 
60.37 d orbital period \citep{williams10}. First, we measured the equivalent 
width ($EW$ hereafter) by integrating the H$\alpha$ emission 
profile in the 64 individual spectra. In addition, the  $FWHM$ and velocity 
shift of the 
line centroid were extracted through simple Gaussian fits. The left panel of 
Figure~\ref{fig2} 
presents the evolution of the H$\alpha$ parameters with time. Because our base
line extends over just 96 days we cannot probe for the presence of a 60.37 d modulation  
through a period search analysis.

The $EW$ is fairly stable during our observations with a mean of $\sim$23 \AA 
~and a smooth increase toward the end of our observing window. A peak is 
detected at HJD 2455725 but, unfortunately, we lack the necessary time coverage 
to check for repeatability with a 60 d period. On the other hand, the $FWHM$ 
and centroid velocity do display significant variability, with two maxima 
elapsed by $\sim$60 d, in good agreement with the photometric modulation of 
\cite{williams10}. The behaviour of the H$\alpha$ line parameters with the 
60.37 d period is presented in the right panel of Figure~\ref{fig2}. 

\begin{figure}
\centering
\includegraphics[angle=0, scale=0.35]{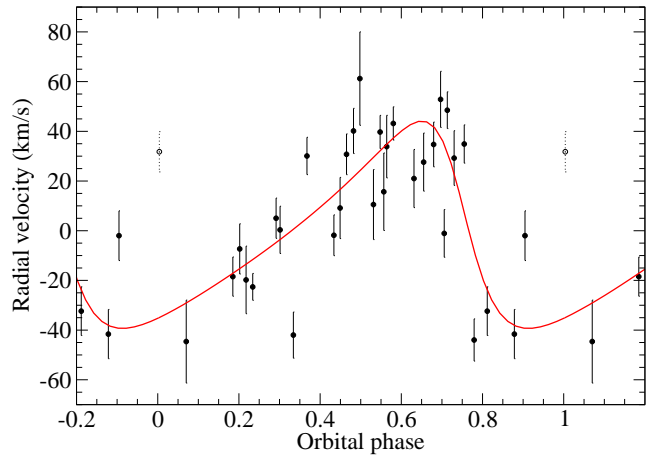}
\caption{Radial velocity curve of MWC~656 folded on the 60.37 d period. 
The best-fitting solution, using an eccentric orbit with $e=0.4$, is overplotted. 
The dotted point (at phase $\sim$0.0) has been masked from the fit. Phase 0 is set 
to HJD 2453243.3. 
}
\label{fig3}
\end{figure}

Next, we attempted to measure radial velocities from the photospheric lines of
the Be star. This is 
complicated by the fact that these lines are extremely broad and, in most 
cases, blended with disc emission lines e.g. the He~{\sc i} lines at 4922 \AA 
~and 5015 \AA ~partly overlap with Fe~{\sc ii} emissions at 4924 \AA ~and 
5018 \AA ~(see Figure~\ref{fig1}).  
After careful comparison of the average spectrum of MWC~656 with the broadened 
template HR 2479, we decided to restrict the radial velocity analysis 
to three spectral windows covering the He~{\sc i} lines 4471 \AA , 4713 \AA 
~and 5048 \AA .
We measured radial velocities by cross-correlating each spectrum of MWC~656 
with a template formed from the average of the entire database. We find that 
this yields better results than using the broadened version of HR 2479 as 
template. 
The radial velocity of the template was determined by fitting a 
Gaussian to the core of the He~{\sc i} 4471 \AA ~line and it was added to the 
velocities obtained from the cross-correlation. 
Figure~\ref{fig3} presents the final radial velocities folded on the 60.37 d 
period of 
\cite{williams10}. The plot reveals a sine-like modulation which hints
for a moderate eccentricity. 

The radial velocity curve was subsequently modelled with an eccentric orbital 
solution using the Spectroscopic Binary Orbit Program (SBOP, 
\citealt{etzel85}). 
The orbital period was fixed to 60.37 d. and phase 0 arbitrarily set to HJD 2453243.3 
which corresponds to the epoch of maximum brightness in the photometric light curve 
\citep{williams10}. Individual points were weighted proportionally to $1/\sigma^2$, 
where $\sigma$ is the radial velocity uncertainty.  
In the solutions we adjusted the following orbital parameters:  
argument of the periastron ($\omega$), systemic velocity ($\gamma$), phase of the periastron 
($\phi_{\rm peri}$) and velocity semiamplitude ($K_{\rm opt}$). 
The solution does not converge if the orbit eccentricity ($e$) is left free. 
Therefore, we attempted several fits fixing the eccentricity by hand and find a minimum 
rms for $e=0.4$. Fitting tests using several orbital configurations indicate that an 
uncertainty of $\pm0.1$ for $e=0.4$ is appropriate. 
Note that we have masked the one discrepant point at phase 0 from the fit. A close look 
to this spectrum shows a flatter He~{\sc i} 4471 \AA ~profile than the rest, possibly due 
to disc emission. In any case, this point has a marginal impact in the final 
solution since it slightly reduces the eccentricity to 0.35 while the remaining 
orbital parameters are virtually unaffected.    
Table \ref{tabrvfit} presents our final best-fitting parameters. 

\begin{table}
\centering
\caption[]{Orbital solution for MWC~656 and MWC~148.}
\label{tabrvfit}
\begin{tabular}{lcc}
\hline
\hline
Parameter                   &           MWC~656          &	     MWC~148	        \\
\hline
$P_{\rm orb}$ (days)        &       60.37 (fixed)        &	   321 (fixed)          \\
$T_0$ (HJD-2,450,000)       &       3243.3 (fixed)       &        4857.5 (fixed)        \\
$e$                         &	    0.4 (fixed) 	 &	  0.83$\pm$0.08         \\
$\omega$ (deg)              &	    71$\pm$23	         &	  129$\pm$17            \\
$\gamma$ (km s$^{-1}$)      &	    $-$2.8$\pm$9.4	 &	  48.3$\pm$8.9	        \\
$\phi_{\rm peri}$           &	    0.74$\pm$0.05	 &	  0.967$\pm$0.008       \\
$K_{\rm opt}$ (km s$^{-1}$) &        41.7$\pm$6.8	 &	  22.0$\pm$5.7          \\
$a_1 \sin i$ (R$_{\odot}$)  &	    45.6$\pm$7.3	 &	  77.6$\pm$25.9         \\
$f(M)$ (M$_{\odot}$)        &    0.35$^{+0.20}_{-0.15}$  &     0.06$^{+0.15}_{-0.05}$   \\
$\sigma$ (km s$^{-1}$)      &		20.3		 &	      12.8	        \\
\hline
\end{tabular}
\end{table}

\section{Analysis of MWC~148}

\begin{figure*}
\centering
\includegraphics[angle=0, width=175mm]{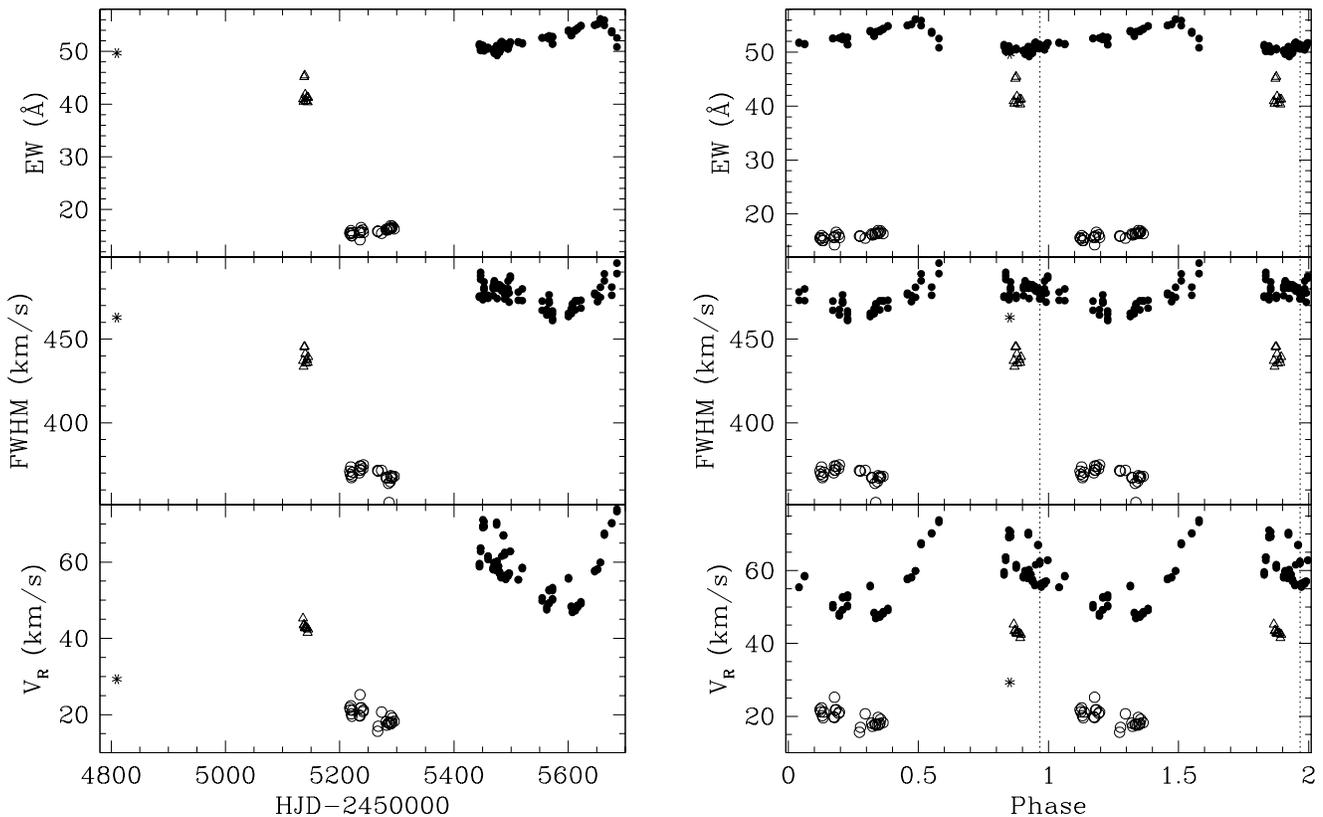}
\caption[]{Left: time evolution of the H$\alpha$ line parameters in MWC~148. Different 
symbols indicate different datasets as follows: asterisk (WHT), open triangles 
(MT) open circles (ST) and filled circles (LT). Note the sinusoidal modulation 
in the LT data with a timescale of $\approx$300 d. Errorbars are not plotted 
because they are always smaller than the symbol size. 
Right: same as in the left panel but folded on the 321 d X-ray period of 
\cite{bongiorno11}. Phase 0 has been set to HJD 2454857.5. Periastron is marked by a
vertical dotted line (see text).
}
\label{fig4}
\end{figure*}

\begin{figure}
\centering
\includegraphics[angle=0, width=84mm]{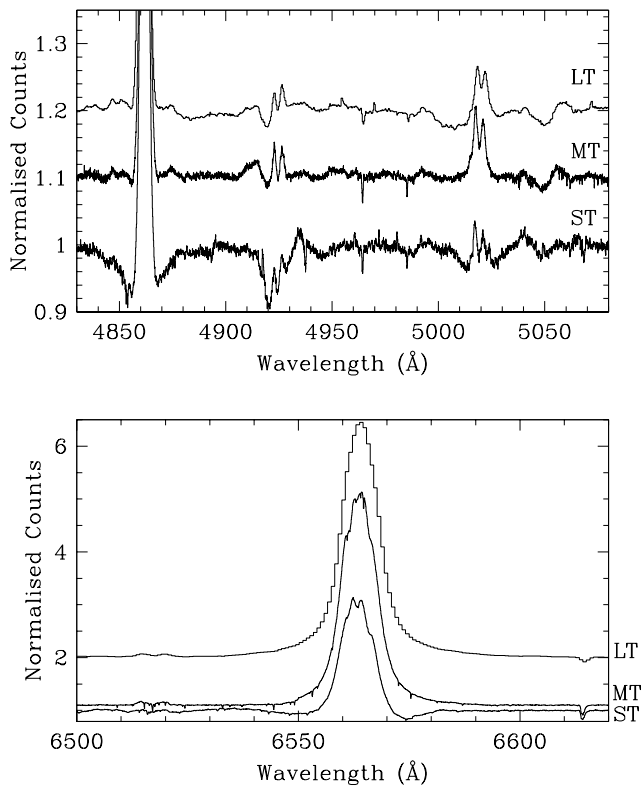}
\caption[]{Average spectra of MWC~148 showing the evolution of several 
circumstellar disc profiles. The emission lines weaken because of the 
appearence of broad absorption throats, especially clear in the ST spectra.
}
\label{fig5}
\end{figure}

In this section we repeat the previous analysis for the star MWC~148. 
The left panel in Figure~\ref{fig4} 
presents the behaviour of the main H$\alpha$ parameters over our baseline of nearly three 
years. The LT data reveal a smooth sinusoidal modulation in all three parameters with a
timescale which, at first glance, seems consistent with the 321 d X-ray period of
\cite{bongiorno11}.  This is better depicted in the right panel of Figure~\ref{fig4}, 
where the H$\alpha$ 
parameters are folded on the 321 d period. The plot also shows that the $EW$s of the 
MT and, in particular, the ST spectra are systematically lower than the rest. For 
comparison, \cite{aragona10} report $EW$=52.3 \AA ~on JD 2454757--92.  
The $FWHM$ and centroid velocities are again consistently lower in the MT and ST spectra. 
A plot of the average H$\alpha$ profiles demonstrates that the line becomes 
weaker and narrower in the MT and ST spectra because of the gradual appearence 
of a broad absorption throat with a full-width-zero-intensity $FWZI\sim$2000 km s$^{-1}$ 
(see bottom panel in Figure~\ref{fig5}). 
This is characteristic of the faint states of Be stars which are likely related to 
episodes of reduced circumstellar envelope \citep{grundstrom06}. Broad absorptions are 
also detected in H$\beta$ and the myriad of Fe~{\sc ii} lines which plague the blue 
spectral range such as Fe~{\sc ii} 4924 \AA ~and 5018 \AA ~(see top panel in 
Figure~\ref{fig5}). These absorptions 
will likely contaminate most of the
photospheric lines. Therefore, we decided to exclude these spectra from the 
study of the radial velocity curve of the Be star.      
   
\begin{figure}
\centering
\includegraphics[angle=0, width=84mm]{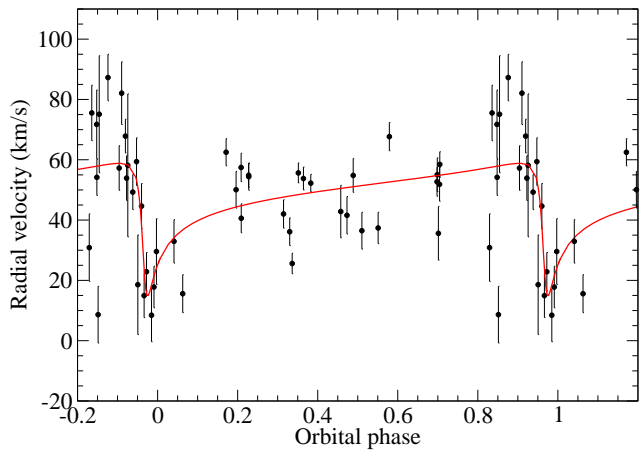}
\caption[]{Radial velocity curve of MWC~148 folded on the 321 d period. 
The best eccentric orbital solution is overplotted. Phase 0 is set to 
HJD 2454857.5.
}
\label{fig6}
\end{figure}

As in MWC~656, radial velocities were obtained by cross-correlating every blue 
INT/WHT/LT spectrum with a template produced from the average of all. 
Prior to this, each spectrum was rebinned into a uniform velocity scale of 23 
km s$^{-1}$ per pixel.  Cross-correlation was performed over spectral windows 
covering the He~{\sc i} lines 4471 \AA , 4713 \AA ~and 5048 \AA. Additional 
regions containing 
lines of 
Si~{\sc iii} (4568 \& 4575 \AA), C~{\sc iii}/O~{\sc ii} (4639--4650 \AA) ~and 
He~{\sc ii} 4686 \AA ~were also included. These absorption features are stronger 
in MWC~148 than in MWC~656 because of its earlier spectral type. The rest 
velocity of the template was again determined through a Gaussian fit to the 
core of the He~{\sc i} 4471 \AA ~line and it was subsequently added to the 
cross-correlation velocities. The radial velocity curve, folded on the 321 d  
period is displayed in Figure~\ref{fig6}. 
In spite of the limited phase coverage and velocity scatter, a sharp 
velocity minimum is detected. The narrowness of this feature prompts for a large 
eccentricity. We have attempted to model this radial velocity curve with the 
SBOP fitting code, fixing the orbital period to 321 d. Following \cite{bongiorno11}, 
we have defined phase 0 as HJD 2454857.5 which was arbitrarily set to the date of the 
first {\it Swift}/XRT observation in their paper.
The best-fitting orbital elements 
are presented in the second column of Table \ref{tabrvfit}.

\section{Discussion}

In this paper we have presented an extensive spectroscopic database of MWC~656 
and MWC~148, the optical counterparts of the candidate $\gamma$-ray binaries 
AGL J2241+4454 and HESS J0632+057 respectively. Our data show evidence for 
long-term modulation in the main H$\alpha$ parameters, consistent with the 
proposed orbital periods of 60.37 d and 321 d respectively. 
These modulations are likely produced by tidal instabilities in the circumstellar 
envelopes triggered by the motion of the compact star in an eccentric orbit. 
Circumstantial support for the orbital nature of the H$\alpha$ variability is 
given by the $EW$ values. The $EW$ of the H$\alpha$ line in Be/X-ray binaries 
provides a good estimate of the size of the circumstellar disc and, because 
this is truncated by the tidal torques of the compact star, a simple relation 
$EW$(H$\alpha$) $\propto P^{4/3}$  is expected \citep{reig11}. It should be 
noted that this correlation holds for the 
maximum $EW$ observed over a long length of time. 
We measure $EW=25$ \AA ~for MWC~656 and 56 \AA ~for MWC~148 which, according 
to the empirical $P_{\rm orb}-EW$(H$\alpha$) diagram in Fig.15 of \cite{reig11}, 
suggest $P_{\rm orb}\approx$ 90 and 250 d respectively. 
Given the scatter of the figure, these are fully consistent with the orbital
periods proposed in the literature. 
 
\subsection{ Masses of the compact stars}

Further, we have detected radial velocity variations in the photospheric lines
of both MWC~656 and MWC~148. Assuming that their orbital periods are 60.37 d and
321 d, we produce phase folded radial velocity curves which allows us to 
constrain their orbital elements for the first time. 
In particular, we determine the mass function of the compact object to be 
$f(M) = M^{3}_{\rm c}\sin^{3} i/\left(M_{\rm c} + 
M_{\rm opt}\right)^2 =0.35^{+0.20}_{-0.15}$~M$_{\odot}$ for MWC~656 and 
$0.06^{+0.15}_{-0.05}$~M$_{\odot}$ for MWC~14. In this equation, $M_{\rm c}$ and 
$M_{\rm opt}$ stand for the masses of the compact and the optical star respectively. 
Unfortunately, no pulsations have been detected from the compact stars yet  
and hence we cannot provide a full solution to the stellar masses and the nature
of the unvisible companions at this point. 
However, some speculations can be made based on our determination of the
rotational broadenings $v \sin i$.  

We measure $v \sin i \sim 346$ km~s$^{-1}$ for MWC~656 and $\sim373$ 
km~s$^{-1}$ for MWC~148. 
A lower limit to the inclination can be derived from the condition
that the optical companion should not exceed 0.9 times the critical rotational 
velocity $v_{\rm crit}$. 
According to  Table 2 in \cite{yudin01}, $v_{\rm crit}\sim$ 565 km~s$^{-1}$ for 
a B0 V which implies $i\ga$47$^{\circ}$ for MWC~148. On the other hand, 
$v_{\rm crit}\sim$ 420 km~s$^{-1}$ for a B3 IV and hence 
$i\ga$66$^{\circ}$ for MWC~656. These estimates obviously assume that the Be 
star's spin axis and the orbital axis are aligned, which may not be true 
considering a possible kick during  
the supernova explosion that forms the compact object.  

In addition, the presence of double peaked emission profiles in our spectra  
also hints for a moderately high inclination. In particular, the Fe~{\sc ii} 
lines are optically thin and their profiles reflect the Keplerian rotation in 
the emitting part of the Be disc \citep{hanuschik96}.
Therefore, we can use their $FWZI$ to estimate the binary inclination 
through $FWZI/2~\times \sin i = (G M_{\rm opt}/R_{\rm opt})^{1/2}$. 
We measure $FWZI$ ($\lambda$5018) $\sim$1300 km s$^{-1}$ in MWC~148 and,
using $M_{\rm opt}\simeq$13.2--19.0~M$_{\odot}$ and 
$R_{\rm opt}\simeq$6.0--9.6~R$_{\odot}$ from \cite{aragona10}, we find 
$i\sim$71--90$^{\circ}$. 
Regarding MWC~656 we measure $FWZI$ ($\lambda$5018) $\sim$1000 km s$^{-1}$ 
which, combined with $M_{\rm opt}\simeq$5.8--9.8~M$_{\odot}$ and 
$R_{\rm opt}\simeq$4.7--8.5~R$_{\odot}$ from \cite{williams10}, yield 
$i\sim$70--76$^{\circ}$.
These estimates, however, should be regarded as mere upper limits because the 
region where the Fe~{\sc ii} lines are produced is unlikely to extend down to 
the surface of the Be star.
 
We also note that the Fe~{\sc ii} profiles do not exhibit deep central absorptions 
below the continuum, characteristic of shell stars. Shell lines are 
thought to be produced by obscuration of the star by the circumstellar disc and 
are only observed at $i\ge80^{\circ}$ \citep{hanuschik96}, which sets a rough 
upper limit to the inclination in both MWC~148 and MWC~656.  
In summary, crude values of $i\simeq$47--80$^{\circ}$ for MWC~148 and 
$i\simeq$67--80$^{\circ}$ for MWC~656 seem the most plausible given the available 
constraints at this point. 
Bringing the constraints to the inclination and the Be stellar masses from 
\cite{williams10} and \cite{aragona10} into the mass function equation results 
in $M_{\rm c}\simeq$2.7--5.5~M$_{\odot}$ for  MWC~656 and 
1.3--7.1~M$_{\odot}$ for MWC~148, although these numbers should be treated with caution 
because of the large
uncertainties involved in this calculation. First, the binary inclinations are loosly
constrained and need to be refined. Further, our orbital solution is based on a limited data 
set which requires confirmation through a more extended baseline. In particular, 
observations of MWC~148 around the periastron at phase 0 are strongly encouraged. 
Unfortunately, the long orbital period coupled to the annual solar cycle means that the 
first opportunity for ground observations will not happen until March 
2015. 

The wide range of masses allowed by our orbital solutions prevent us from deciding 
whether the compact objects in MWC~148 and MWC~656 are neutron stars or black holes. 
However, their long orbital periods place them among the widest 
$\gamma$-ray binaries, only after PSR B1259--63, and this offers an opportunity to 
probe the nature of the 
compact star through the detection of radio pulsations. Previous attempts in 
LS~5039 and LS~I~+61~303 have failed presumably due to strong free-free 
absorption by the dense stellar winds (see McSwain et al. 2011 and references therein). 
In principle, the detection of radio pulses might be possible in both MWC~148 and MWC~656 
around apastron phases due to their wider orbits.
X-ray pulses are also expected to arise through synchrotron emission 
in the magnetosphere of a rotationally-powered pulsar or from an accreting neutron star.     
Deep searches for X-ray pulsations have been performed in both  LS~5039 and LS~I~+61~303 
with null results (See Rea et al 2010, 2011 and included references). This implies than 
the putative pulsars are spinning faster than $\sim$5.6ms, the pulsar beam is pointing away 
from our line of sight or X-ray pulsations are restricted to a limited range of orbital phases. 
We note that \cite{rea11a} could not either find 
pulsations in MWC~148 using $Chandra$ data during one outburst nor in 
$XMM-Newton$ data during quiescence. 
These observations had a time resolution of only 199.2 ms and, according to our ephemeris, 
were performed at orbital phases 0.33 and 0.46, respectively.  
Therefore, the $XMM-Newton$ data coincide with the apastron passage, where 
Compton scattering is lower and hence more favorable for the detection of X-ray pulses. 
More higher time resolution observations of 
both MWC~656 and MWC~148 at different orbital phases 
are urgently required to test the nature of their compact stars.

\subsection{Orbital variability}

\begin{figure*}
\centering
\includegraphics[angle=0, width=175mm]{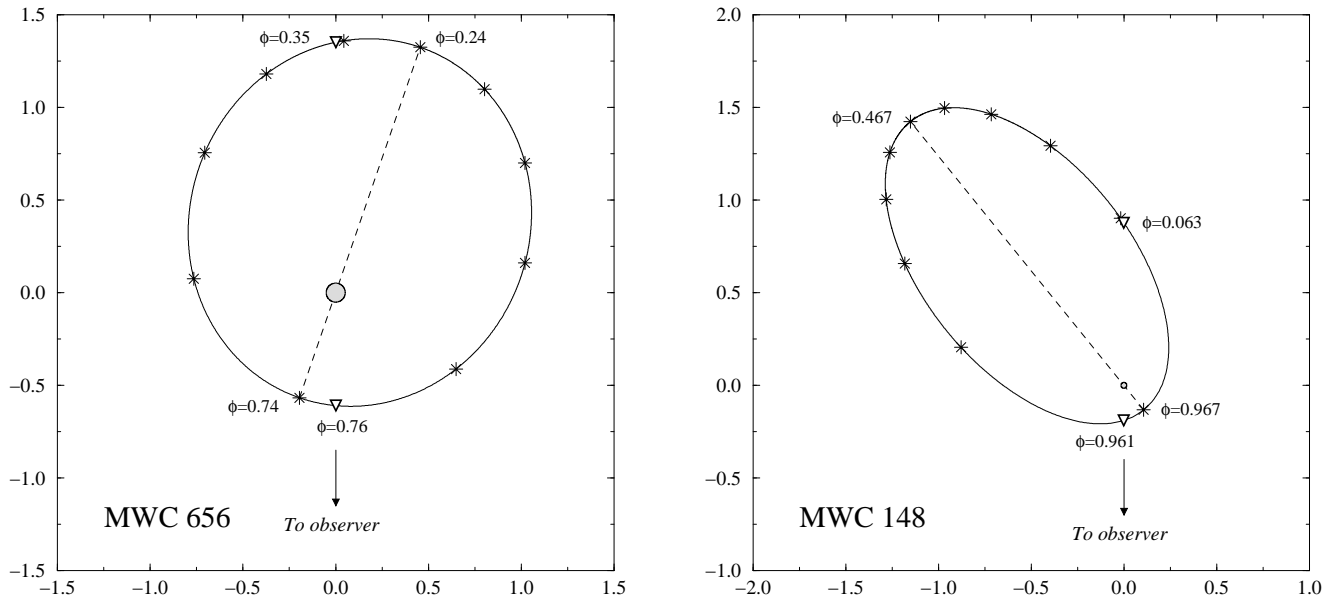}
\caption[]{Left:relative orbit of the compact object around the optical star in MWC~656 as
seen from above the orbital plane. Phases of periastron and apastron are indicated and joined 
by a dashed line which depicts the major axis of the orbit. Stars mark 0.1-phase intervals from 
periastron whereas open triangles indicate the phases of the inferior conjunction and superior 
conjunction of the compact star. 
The coordinates are in units of the orbital semimajor axis.  Right: same for MWC~148. 
Note that the semimajor axis is 0.64 AU in MWC 656 and 2.38 AU in MWC 148
}
\label{fig7}
\end{figure*}

Aside from the stellar masses, our orbital solutions do provide interesting information 
on the geometry of the $\gamma$-ray binaries and properties of the observed X-ray/VHE 
radiation. Figure~\ref{fig7} presents the relative motion of the compact star around the 
optical 
companion for both MWC~656 and MWC~148 as seen from above i.e. $i=0^{\circ}$. 
The figure was produced using the stellar masses and radii from \cite{williams10} and 
\cite{aragona10} and assuming a 1.4 M$_{\odot}$ compact star. 
Our ephemeris imply that the X-ray outbursts in MWC~148 are delayed by $\sim$100 d or 0.3 
orbital phases with respect to the periastron passage. This is remarkably similar to 
LS~I~+61~303 where X-ray outbursts are observed $\sim$0.2 orbital phases passed the 
closest approach of the stellar components \citep{casares05a}.  

Regarding MWC~656, the maximum of the optical modulation takes 
place $\sim$0.25 orbits after periastron. A similar phase shift of the 
photometric maximum is seen in LS~I~+61~303 
\citep{mendelson89, paredes94}. 
In addition, the right frame in Figure~\ref{fig2} shows that 
the $EW$ of the H$\alpha$ line peaks $\sim$0.3 orbital phases passed
periastron. This behaviour 
seems consistent with an scenario where the modulation in both the continuum 
flux and the (opticaly thick) H$\alpha$ flux is caused by the visibility of 
a one-arm tidal wave in the circumstellar disc, triggered by the close passage 
of the neutron star. For instance, \cite{okazaki02} present SPH simulations of a 
Be/X-ray binary with $P_{\rm orb}$=24.3 d and $e=0.34$ which is not too far from 
the parameters that we derive for MWC~656. Their simulations (see Figures 10 and 
11 in the paper) show that a spiral density wave, excited by the periastron passage, 
is fully developed around phase $\sim$0.25 past periastron. 
Furthermore, \cite{williams10} observed a rapid variability of the $V/R$ ratio of the 
H$\gamma$ emission and suggested it could be associated with the changing 
tidal effects of the companion on the disk, which is specially strong near periastron. 
Indeed, these observations correspond to orbital phases $0.82\pm0.02$ and 
$0.84\pm0.02$ which according to our ephemeris are just $\sim$0.08 and 
$\sim$0.10 in phase after periastron.  
At phase $\sim$0.8 we also observe a rapid transition of the H$\alpha$ centroid from 
negative to positive velocities, coincident with a 
peak in the $FWHM$ (see right frame in Figure~\ref{fig2}). 

In view of the new orbital ephemeris of MWC~656 it is also interesting to put 
in context the multiwavelength observations of the source. First of all, the {\it AGILE} 
detection between 25 July 2010 at 01:00~UT and 26 July 2010 at 23:30~UT 
\citep{lucarelli10}, corresponding to an orbital phase range of 
0.77--0.80, took place just after (or even during) periastron. A maximum 
of the GeV emission takes place during periastron in LS~5039 
\citep{abdo09_ls5039} and soon after it in LS~I~+61~303 \citep{abdo09_lsi}. 
The fact that the {\it AGILE} flare in MWC~656 happens very close to the inferior 
conjunction of the compact object ($w=71\pm23$~deg), coupled with its high orbital
inclination, may explain the weak GeV emission through a  
reduction in the Compton scattering cross-section for large angles \citep{khangulyan08}.
All these facts reinforce the association between the Be star MWC~656 and 
the gamma-ray source AGL~J2241+4454 and, even with the lack of $\gamma$-ray 
detections aside from the flare seen by {\it AGILE} during a single 
periastron passage, give strong support to the idea that MWC~656 is a new 
$\gamma$-ray binary.

On the other hand, MWC~148 has only shown significant VHE radiation simultaneous to 
the X-ray flare episodes which occur $\sim$0.3 phases past periastron. Indeed, 
according to our ephemeris, the $VERITAS$ upper limits \citep{acciari09} were 
obtained during the phase intervals 0.6--0.72, 0.86--0.93 and 0--0.01, corresponding to 
low levels of quiescent X-ray emission \citep{bongiorno11}. 
Conversely, the H.E.S.S. discovery observations occured at phases 0.32 and 0.36--0.46, 
coincident with periods of X-ray flares. Further detections of MWC~148 at VHE during 
X-ray active phases have been reported by \cite{ong11} and \cite{mariotti11}.     

\subsection{The class of $\gamma$-ray binaries}

\begin{figure}
\centering
\includegraphics[angle=-90, width=85mm]{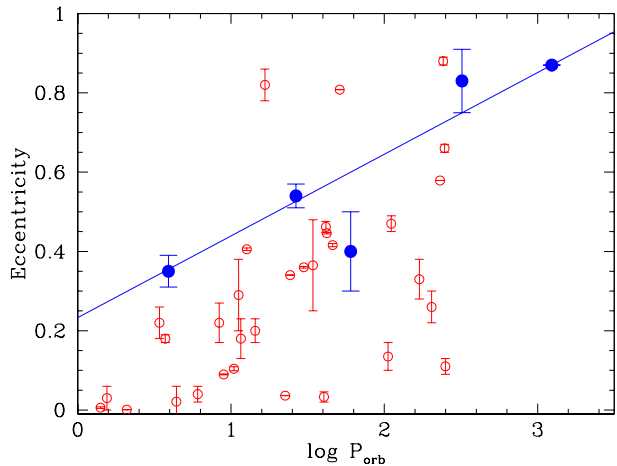}
\caption[]{Eccentricity versus orbital period for $\gamma$-ray binaries (filled
circles) and high mass X-ray binaries from the catalogue of Liu et al. 2007
(open circles). A linear fit to the $\gamma$-ray binaries is overplotted.
}
\label{fig8}
\end{figure}

The family of $\gamma$-ray binaries is growing fast with the recent discovery 
of AGL J2241+4454, HESS J0632+057.1 and 1FGL J1018.6$-$5856. These add up to the 
group of  "classic" $\gamma$-ray binaries PSR B1259$-$63, LS 5039 and LS~I~+61~303. 
Table \ref{tab4} summarizes some of their main physical parameters. They all 
contain late O-Be stars with strong winds and stellar radii in the range 
$\sim$7--10~R$_{\odot}$. 
The table shows a possible correlation between the eccentricity and the 
orbital period and this is illustrated in figure~\ref{fig8}. A least-squared 
linear fit yields:  

\begin{equation}
e = 0.206 (19) \times \log P_{\rm orb} + 0.233 (60)
\end{equation}

\noindent
For comparison, Figure~\ref{fig8} also presents 35 high-mass X-ray binaries (HMXBs) with 
X-ray pulsars from \cite{liu06} and \cite{martin09}. 
HMXBs tend to lie below the linear regresion of  
$\gamma$-ray binaries, i.e. for a given orbital period they have lower eccentricities. 
The only exceptions are 4U 1850$-$03, 0535$-$668 and the SMC binary J0045$-$7319, all 
with $e>0.8$. In particular, the latter two have $P_{\rm spin}<1$ s 
and hence a high spin-down power which 
makes them 
potential candidates for $\gamma$-ray emission. 

The reason behind the $P_{\rm orb} -e$ correlation in $\gamma$-ray binaries stems from the fact 
that small stellar separations 
are required to trigger VHE radiation and this can only happen for long orbital 
periods if the eccentricity is large. Conversely, for short 
orbital periods high eccentricities would imply that the compact object 
passes through the stellar atmosphere of the companion, clearly an 
unstable situation. 
The separation of the two stars at periastron $a(1-e)$ and apastron $a(1+e)$ is 
also listed in Table \ref{tab4}, 
assuming that the compact star is a 1.4 M$_{\odot}$ neutron 
star. As expected, the most compact binaries LS 5039, 1FGL J1018.6$-$5856 and 
LS~I~+61~303 show persistent VHE and/or HE radiation modulated with their orbital 
periods whereas only transient radiation is detected in PSR B1259$-$63 around periastron. 
This seems to imply that stellar separations $\la$1 a.u. or $\la$25 $R_{\rm opt}$ are 
required for the efficient production and modulation of the HE/VHE radiation but the 
picture is certaintly more complex. For instance, 
MWC~656 has been detected  up to GeV energies only once during a flare episode at 
periastron whereas MWC~148 is regularly detected at TeV $\sim$0.3 
orbital phases past periastron. Aside from sensitivity limitations of different 
satellite/telescopes, other processes such as photon-photon annihilation and the 
subsequent production of electron-positron pairs are likely responsible for the wide 
phenomenology observed. In particular, the binary inclination and argument of periastron 
are important parameters for the attenuation of the VHE radiation and modelling the 
observed amplitudes \citep{dubus06}. 
 
\begin{figure}
\centering
\includegraphics[angle=-90, width=85mm]{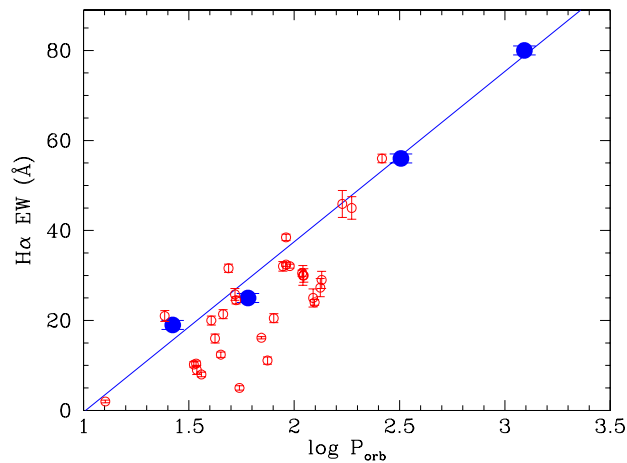}
\caption[]{ Maximum $EW$ of the H$\alpha$ emission line versus orbital period for
$\gamma$-ray binaries (filled circles) and Be/X-ray binaries (open circles). The latter include 
Milky Way and SMC binaries from Reig (2011). A linear fit to the $\gamma$-ray binaries 
is overplotted. Note that LS 5039 is not included because the primary is not a Be star.  
}
\label{fig9}
\end{figure}

It is well established that decretion discs in Be/X-ray binaries are truncated by the tidal 
torque of the neutron star \citep{negueruela01}. Since the $EW$ of the H$\alpha$ emission 
is a good proxy of the disc size this results in the well known
correlation between $P_{\rm orb}$ and the maximum observed $EW (H\alpha)$ \citep{reig97, reig11}. 
Most $\gamma$-ray binaries contain Be primaries and it is interesting to test whether a 
similar correlation holds. Table \ref{tab4} lists the range of $EW (H\alpha)$ values reported 
in literature for the $\gamma$-ray binaries. 
Significant long-term and orbital variability is typically seen in Be/X-ray binaries 
and the same is observed in the $\gamma$-ray binaries. In this paper we detect a maximum $EW$ 
for MWC~656 at $\sim$ 0.3 phases past periastron, in agreement with what is observed in LS~I~+61~303 
\citep{mcswain10}. On the other hand, the maximum $EW$ for MWC~148 occurs at periastron. 
In any case, only the maximum $EW$ values are driven by the disc truncation radius.  
Figure~\ref{fig9} plots the maximum $EW (H\alpha)$ measured in $\gamma$-ray binaries with Be 
primaries versus $P_{\rm orb}$ and we note that they follow the linear regresion:  

\begin{equation}
EW = 37.8 (5.0) \times \log P_{\rm orb} -38.3 (11.4)
\end{equation}

\noindent
For comparison, Figure~\ref{fig9} also shows $EW$s of Be/X-ray binaries from \cite{reig11}.  
As in the case of the eccentricity, we observe that $\gamma$-ray binaries define the higher envelope of the
distribution i.e. for a given orbital period Be/X-ray binaries tend to have smaller decretion discs than 
$\gamma$-ray binaries. This is somehow surprising in   
a millisecond pulsar scenario because it implies that the relativistic pulsar wind does not "erode" 
significantly the circumstellar disc even during periastron passage. 
However, it is also true that, given the large eccentricities of $\gamma$-ray binaries, 
they only spend a short fraction of time close to the Be star, perhaps preventing efficient 
truncation.  Interestingly, equations 1 and 2 provide a new tool for a rough estimate of two fundamental 
parameters (such as $P_{\rm orb}$ and $e$) in Be/$\gamma$-ray binaries simply from the 
$EW$ of the H$\alpha$ emission. This can be tested with future observations of newly discovered 
$\gamma$-ray binaries.

\begin{table*}
\centering
\begin{minipage}[t]{15cm}
\caption[]{The family of $\gamma$-ray Binaries sorted by orbital period}
\label{tab4}
\renewcommand{\footnoterule}{}  
\begin{tabular}{l@{~~}l@{~~}l@{~~}c@{~~}c@{~~}c@{~~}c@{~~}c@{~~}c@{~~}c@{~~}c@{~~}c@{~~}c@{~~}}
\hline
\hline
Name  & $\gamma$-ray  & Spectral  &   $M_{\rm opt}$ &   $R_{\rm opt}$  & $i$ & $P_{\rm orb}$  & $e$ &   $a(1-e)$ &  $a(1+e)$ &  $EW$ & $d$ & Reference~\footnote{
(1) Casares et al. 2005b. (2) Corbet et al. 2011. (3) Napoli et al. 2011. (4) Aragona et al. 2009. (5) Zamanov et al. 1999. 
(6) Casares et al. 2005a. (7) McSwain et al. 2010. (8) Williams et al. 2010. (9) Aragona et al. 2010.    
(10) Negueruela et al. 2011. (11) Johnston et al. 1994. (12) Grundstrom et al., private communication}  \\
     & Activity  & Type      &  (M$_{\odot}$) &  (R$_{\odot}$) &  (deg)&   (days)     &   &    (AU) &    (AU) &  (\AA ) &   (kpc) &     \\
\hline
LS~5039             & HE, VHE   &  O6.5 V((f)) &   20--26    &  9--10    & 13--64      &   3.91  & 0.35  &  0.09  & 0.19  & $-$2.8  & 2.5       & (1) \\
1FGL J1018.6$-$5856 & HE        &  O6 V((f))   &  $\sim$37   & $\sim$10  &  --         &  16.58  &  --   &   --   &  --   & --      & $\sim$5.4 & (2), (3) \\
LS~I~+61~303        & HE, VHE   &  B0 Ve       &   10--15    & $\sim$7   & 10--60      &  26.50  &  0.54 &  0.19  & 0.64  & 8--19   & 1.9       & (4), (5), (6), (7) \\
MWC~656             & HE        &  B3 IVne+sh  &   6--10     &  5--9     & 67--80      &  60.37  &  0.40 &  0.38  & 0.89  & 19--25  & $\sim$2.6 & (8), this paper \\
MWC~148             & ~~~~~~VHE &  B0 Vpe      &   13--19    &  6--10    & 47--80      &  321    &  0.83 &  0.40  & 4.35  & 48--56  & $\sim$1.4 & (9), this paper \\
PSR B1259$-$63      & HE, VHE   &  O9.5 Ve     &   31        &  9        & 19--31      & 1236.79 &  0.87 &  0.93  & 13.44 & 40--80  & 2.3       & (10), (11), (12)	   \\
\hline
\end{tabular}
\end{minipage} 
\end{table*}
\normalsize                

\section{Summary}

We have reported optical spectroscopy of MWC~656 and MWC~148, the optical counterparts of the 
$\gamma$-ray sources AGL~J2241+4454 and HESS~J0632+057 respectively. Our data show that the main 
H$\alpha$ parameters are modulated with the 60.37 d optical photometric period in MWC~656 and    
with the 321 d X-ray period in MWC~148. This is likely produced by the visibility of a tidal wave 
in the Be disc, triggered by the close passage of a compact star. 
In addition, we present radial velocity curves for the two stars and the first constraints 
to the binary parameters. Both stars display similarities with LS~I~+61~303, where 
photometric maxima and X-ray outburst are delayed by $\sim$0.3 phases past periastron. 
When compared to other $\gamma$-ray binaries, we find that the eccentricity is correlated with 
the orbital period. This is explained by the small stellar separations needed for triggering 
the $\gamma$-ray activity. The maximum $EW$ of the H$\alpha$ emission is also correlated 
with orbital period. This is reminiscent of Be/X-ray binaries and suggest that the size of  
circumstellar discs in $\gamma$-ray binaries is tidally truncated by the compact companion.

\section{Acknowledgments}

We thank D. Steeghs and L. van Spaandonk for kindly taking the WHT spectra. We are also greatful 
to E. Grundstrom for sharing her EW measurements of PSR B1259$-$63 with us and to P. Reig 
for the EW data of Be X-ray binaries. 
Based on observations made with the INT and WHT operated on the 
island of La Palma by the Isaac Newton Group in 
the Spanish Observatorio del  Roque de Los Muchachos of the Instituto
de Astrof\'\i{}sica de Canarias (IAC). The Liverpool Telescope is 
operated on the island of La Palma by Liverpool John Moores University 
in the Spanish Observatorio del Roque de los Muchachos of the Instituto 
de Astrofisica de Canarias with financial support from the UK Science 
and Technology Facilities Council. Also based on observations made 
with the Mercator operated on the island of La Palma by the Univ. of 
Leuven and the Obs. of Geneva 
in the Spanish Observatorio del  Roque de Los Muchachos of the IAC. 
This research has been supported by the Spanish Ministerio de Ciencia e Innovaci\'on 
(MICINN) under grants AYA2010-18080, AYA2010-21782-C03-01, 
AYA2010-21697-C05-05 and FPA2010-22056-C06-02.
M.R. acknowledges financial support from MICINN and European Social Funds
through a \emph{Ram\'on y Cajal} fellowship.
J.M.P. acknowledges financial support from ICREA Academia. 
Partly funded by the Spanish MEC under the Consolider-Ingenio 2010 Program 
grant CSD2006-00070: ''First science with the GTC'' 
(http://www.iac.es/consolider-ingenio-gtc/).
MOLLY software developed by T. R. Marsh is gratefully acknowledged. 

{}


\begin{thebibliography}{}

\bibitem[Abt et al.(2002)]{abt02}
Abt H.A., Levato H., Grosso M., 2002, ApJ, 573, 359

\bibitem[Abdo et al.(2009a)]{abdo09_ls5039}
Abdo, A.~A., Ackermann, M., Ajello, M., et al., 2009a, ApJ, 706, L56 

\bibitem[Abdo et al.(2009b)]{abdo09_lsi}
Abdo, A.~A., Ackermann, M., Ajello, M., et al., 2009b, ApJ, 701, L123

\bibitem[Acciari et al.(2009)]{acciari09}
Acciari, V.~A., et al., 2009, ApJ, 698, L97

\bibitem[Aharonian et al.(2007)]{aharonian07}
Aharonian F.A. et al., 2007, A\&A, 469, L1

\bibitem[Aragona et al.(2009)]{aragona09}
Aragona C. et al., 2009, ApJ, 698, 514

\bibitem[Aragona et al.(2010)]{aragona10}
Aragona C., McSwain M.V., De Becker M, 2010, ApJ, 724, 306

\bibitem[Bongiorno et al.(2011)]{bongiorno11}
Bongiorno S.D. et al., 2011, ApJ, 737, L11

\bibitem[Dhawan et al.(2006)]{dhawan06}
Dhawan V., Mioduszewski A., Rupen M., 2006, PoS, Procceedings of the VI 
Microquasar Workshop: Microquasars and Beyond, ed. T. Belloni, 52.1

\bibitem[Dubus(2006)]{dubus06}
Dubus G., 2006, A\&A, 451, 9

\bibitem[Casares et al.(2005a)]{casares05a}
Casares J., Ribas I., Paredes J.M., Mart\'\i{} J., Allende Prieto C., 2005a, 
MNRAS, 360, 1105

\bibitem[Casares et al.(2005b)]{casares05b}
Casares J., Rib\'o M., Ribas I., Paredes J.M., Mart\'\i{} J., Herrero A., 2005b, 
MNRAS, 364, 899

\bibitem[Casares et al.(2011)]{casares11}
Casares J. et al., 2011, in "High-Energy Emission from Pulsars and their Systems",  
Astrophysics and Space Science Proceedings, ed. N. Rea and D.F. Torres,
Springer-Verlag, p. 559

\bibitem[Corbet et al. (2011)]{corbet11}
Corbet R.H.D. et al., 2011, ATel, 3221

\bibitem[\protect\citeauthoryear{Etzel}{1985}]{etzel85}
Etzel, P. B., 1985, "SBOP - Spectroscopic Binary Orbit Program", Program's Manual

\bibitem[Falcone et al.(2010)]{falcone10}
Falcone A.D. et al., 2010, ApJ, 708, L52

\bibitem[Finger et al.(1999)]{finger99}
Finger M.H. et al., 1999, ApJ, 517, 449

\bibitem[Gray(1992)]{gray92}
Gray D.F., 1992, The Observation and Analysis of Stellar Photospheres. 
CUP, Cambridge

\bibitem[Guti\'errez-Soto et al.(2007)]{guti07}
Guti\'errez-Soto J., Fabregat J, Suso J., Lanzara M., Garrido R., 
Hubert A.-M., Floquet M., 2007, A\&A, 476, 927

\bibitem[Grundstrom \& Gies (2006)]{grundstrom06}
Grundstrom E.D., Gies D.R., 2006, ApJ, 651, L53

\bibitem[Hanuschik(1996)]{hanuschik96}
Hanuschik R.W., Hummel W., Sutorius E., Dietle O., Thimm G., 1996, A\&AS, 116, 309

\bibitem[Hinton \& Hofmann (2009)]{hinton09a}
Hinton J.A., Hofmann W., 2009, ARA\&A, 47, 523

\bibitem[Hinton et al.(2009)]{hinton09b}
Hinton J.A.. et al., 2009, ApJ, 690, L101

\bibitem[Horne(1986)]{horne86}
Horne K., 1986, PASP, 98, 609

\bibitem[Johnston et al. (1994)]{johnston94}
Johnston S., Manchester R.N., Lyne A.G., Nicastro L., Spyromilio J., 1994, MNRAS, 268, 430

\bibitem[Khangulyan et al. (2008)]{khangulyan08}
Khangulyan D., Aharonian F., Bosch-Ramon V., 2008, MNRAS, 383, 467

\bibitem[Liu et al. (2006)]{liu06}
Liu Q.Z., van Paradijs J., van den Heuvel E.P.J., 2006, A\&A, 455, 1165

\bibitem[Lucarelli et al. (2010)]{lucarelli10}
Lucarelli F. et al., 2010, ATel, 2761

\bibitem[Mariotti et al.(2011)]{mariotti11}
Mariotti, M., et~al., 2011, ATel, 3161

\bibitem[Marsh et al.(1994)Marsh, Robinson \& Wood]{marsh94}
Marsh T.R., Robinson E.L., Wood J.H., 1994, MNRAS, 266, 137

\bibitem[Martin et al.(2009)Martin, Tout \& Pringle]{martin09}
Martin R.G., Tout C.A., Pringle J.E., 2009, MNRAS, 397, 1563

\bibitem[McSwain et al.(2010)]{mcswain10}
McSwain M.V., Grundstrom E.D., Gies D.R., Ray P.S., 2010, ApJ, 724, 379

\bibitem[McSwain et al.(2011)]{mcswain11}
McSwain M.V., Ray P.S., Ransom S.M., Roberts M.S.E., Dougherty S.M., Pooley G.G., 
2011, ApJ, 738, 105

\bibitem[Mendelson \& Mazeh(1989)]{mendelson89}
Mendelson, H., \& Mazeh, T., 1989, MNRAS, 239, 733

\bibitem[Mold\'on et al. (2011b)]{moldon11b}
Mold\'on J., Johnston S., Rib\'o M., Paredes J.M., Deller A.T., 2011b, ApJ, 732, L10 

\bibitem[Mold\'on et al. (2011a)]{moldon11a}
Mold\'on J., Rib\'o M., Paredes J.M., 2011a, A\&A, 533, L7 

\bibitem[Napoli et al. (2011)]{napoli11}
Napoli, V.J., McSwain M.V., Boyer A.N.M., Roettenbacher R.M., 2011, PASP, 123,
1262 
 
\bibitem[Negueruela \& Okazaki (2001)]{negueruela01}
Negueruela, I., Okazaki A.T., 2001, A\&A, 369, 108 

\bibitem[Negueruela et al. (2011)]{negueruela11}
Negueruela, I., Rib\'o M., Herrero A., Lorenzo J., Khangulyan D., Aharonian F.A., 2011, ApJ, 732, L11 

\bibitem[Ong et al.(2011)]{ong11}
Ong, R., et~al., 2011, ATel, 3153

\bibitem[Okazaki et al.(2002)]{okazaki02}
Okazaki A.T., Bate M.R., Ogilvie G.I., Pringle J.E., 2002, MNRAS, 337, 967 

\bibitem[Paredes et al.(1994)]{paredes94}
Paredes, J.~M., Marziani, P., Mart\'{\i}, J., et~al. 1994, A\&A, 288, 519

\bibitem[Paredes (2011)]{paredes11}
Paredes J.M., 2011, Il Nuovo Cimento C - Colloquia on physics, arXiv1101.4843 

\bibitem[Porter \& Rivinius(2003)]{porriv03}
Porter J.M., Rivinius Th., 2003, PASP, 115, 1153

\bibitem[Rea et al. (2010)]{rea10}
Rea N., Torres D.F., van der Klis M., Jonker P.G., M\'endez M., Sierpowska-Bartosik A., 
2010, MNRAS, 405, 2206	 

\bibitem[Rea et al. (2010)]{rea11a}
Rea N., Torres D.F., 2011, ApJ, 737, L12	 

\bibitem[Rea et al. (2011)]{rea11b}
Rea N. et al., 2011, MNRAS, 416, 1514	 

\bibitem[Reig et al. (1997)]{reig97}
Reig P., Fabregat J, Coe M.J., 1997, A\&A, 322, 193	 

\bibitem[Reig (2011)]{reig11}
Reig P., 2011, Ap\&SS, 332, 1	 

\bibitem[Rib\'o et al. (2008)]{ribo08}
Rib\'o M., Paredes J.M., Mold\'on J., Mart\'\i{} J., Massi M.,  2008, A\&A, 481, 17 

\bibitem[Skilton et al.(2009)]{skilton09}
Skilton J.L. et al., 2009, MNRAS, 399, 317 

\bibitem[Townsend et al.(2007)]{townsend07}
Townsend R.H.D., Owocki S.P., Ud-Doula A., 2007, MNRAS, 382, 139 

\bibitem[Williams et al.(2010)]{williams10}
Williams S.J. et al., 2010, ApJ, 723, L93 

\bibitem[Yudin (2001)]{yudin01}
Yudin R.V., 2001, A\&A, 368, 912

\bibitem[Zamanov et al. (1999)]{zamanov99}
Zamanov R.K., Mart\'\i~ J., Paredes J.M., Fabregat J., Rib\'o M., Tarasov A.E., 1999, A\&A, 351, 543

\end{thebibliography}
\end{document}